\documentclass[11pt,twoside]{article} 
\usepackage{acta-info,euler}
\usepackage{tikz,pgf}

\newcommand{\vol}{\textbf{3,} 1 (2011) 127--136}  
\newcommand{\amss}{\amssubj{68R15}} 

\newcommand{\CRs}{\CRsubj{G.2.1, F.2.2}}
\newcommand{\keyw}{\keywords{word complexity, scattered subword, $d$-complexity, super-$d$-complexity}}

\newcommand{\key}[1]{\textbf{#1}}
\newcommand{\tsc}[1]{\textrm{\textsc{#1}}}

\newcommand{\alge}{%
\begin{tabbing}%
99 \= xxx\=xxx\=xxx\=xxx\=xxx\=xxx\=xxx\=xxx\=xxx\=xxx\=xxx \+ \kill 
}

\newcommand{\algeN}{%
\begin{tabbing}%
\hspace*{1pt}999\=xxx\=xxx\=xxx\=xxx\=xxx\=xxx\=xxx\=xxx\=xxx\=xxx\=xxx \+ \kill  
}

\newcommand{\algv}{%
\end{tabbing}
}

\newenvironment{algN}[1]{%
\vspace{4mm}  
\vbox\bgroup\noindent\tsc{#1}%
\vspace*{-2mm}  
\algeN}
{
\algv\egroup
\vspace{0mm}  
}

\begin{document}
\title{On scattered subword complexity}
\maketitle

\oneauthor{%
\href{http://www.ms.sapientia.ro/~kasa}{Zolt\'an K\'ASA}
}{
\href{http://www.emte.ro}{Sapientia Hungarian University of Transylvania}\\ 
\href{http://www.ms.sapientia.ro/}{Department of Mathematics and Informatics,} \\Tg. Mure\c s, Romania 
}{
\href{mailto:kasa@ms.sapientia.ro}{kasa@ms.sapientia.ro}
} 

\short{%
Z. K\'asa
}{%
On scattered subword complexity
} 

\begin{abstract} Special scattered subwords, in which the gaps are of length from a given set, are defined. The  scattered subword complexity, which is the number of such scattered subwords, is computed for rainbow words.  

\end{abstract}

\section{Introduction} 
Sequences of characters called \emph{words} or \emph{strings} are widely studied in combinatorics, and used in  
various fields of sciences (e.g.  chemistry, physics, social sciences, biology \cite{fiz, elzinga, elzinga2, bio} etc.).
The elements of a word  are called \emph{letters}. A contiguous part of a word (obtained by erasing a prefix or/and a suffix) is a \emph{subword} or \emph{factor}. If we erase arbitrary letters from a word, what is obtained is a \emph{scattered subword}. Special scattered subwords, in which the consecutive letters are at distance at most $d$ $(d\ge 1)$ in the original word, are called $d$-\emph{subwords} \cite{ivanyi, kasa}. In \cite{kasa1} the \emph{super}-$d$-\emph{subword} is defined, in which case the distances are of length at least $d$. The super-$d$-complexity, as the number of such subwords, is computed for rainbow words (words with pairwise different letters).

In this paper we define special scattered subwords, for which the distance in the original word of length $n$ between two letters which will be consecutive in the subword, is taken from a subset of $\{1,2, \ldots, n-1\}$.
 
The \emph{complexity of a word} is defined as the number of all its different subwords. Similar definitions are for \emph{$d$-complexity}, \emph{super-$d$-complexity} and  \emph{scattered subword complexity}. 

The scattered subword complexity is computed in the special case of rainbow words. The idea of using scattered words with gaps of length between two given values is from J\'ozsef Bukor \cite{bukor}.

Another point of view of scattered complexity in the case of non-primitive words is given is \cite{fazekas}.

\section{Definitions}
Let $\Sigma$ be an alphabet, $\Sigma^n$, as usually, the set of all  words of length $n$ over $\Sigma$, and  $\Sigma^*$ the set of all finite word over $\Sigma$.

\begin{definition}
Let  $n$ and $s$ be positive integers, $M\subseteq \{1,2,\ldots, n-1\}$ and $u=x_1x_2\ldots x_n\in \Sigma^n$. An {$M$-subword} of length $s$ of $u$ is defined as $v=x_{i_1}x_{i_2}\ldots x_{i_s}$ where 

$i_1\ge 1$, 

$i_{j+1}-i_j\in M$  for $j=1,2,\ldots, s-1$,

$i_s\le n.$
\end{definition}
 
\begin{definition}
The number of $M$-subwords of a word $u$ for a given set $M$ is the scattered subword complexity, simply $M$-complexity.
\end{definition}
 
The $M$-subword in the case of $M=\{1,2, \ldots, d\}$  is the $d$-\emph{subword} defined in \cite{ivanyi}, while 
in the case of $M=\{d, d+1, \ldots, n-1\}$ is the \emph{super}-$d$-\emph{complexity} defined in \cite{kasa1}.

\noindent\textbf{Examples.}  The word $abcd$ has 11 $\{1,3\}$-subwords: $a$,  $ab$, $abc$, $abcd$, $ad$, $b$, $bc$, $bcd$, $c$, $cd$,  $d$.  The $\{2,3\ldots, n-1\}$-subwords of the word $abcdef$  are the following: $a$, $ac$, $ad$, $ae$, $af$, $ace$, $acf$, $adf$, $b$, $bd$, $be$, $bf$, $bdf$, $c$, $ce$, $cf$, $d$,  $df$, $e$, $f$.

\medskip
Hereinafter instead of $\{d_1,d_1+1, \ldots, d_2-1, d_2\}$-subword we will use the simple notation $(d_1,d_2)$-subword.

\section{Computing the scattered complexity for rainbow words}
Words with pairwise different letters are called \emph{rainbow words}. 
The $M$-comple\-xity of a rainbow word of length $n$ does not depend on what letters it contains, and is denoted by $K(n,M)$.

Let us recall two results for special scattered words, as $d$-subwords and super-$d$-subwords.

For a rainbow word of length $n$ the super-$d$-compexity \cite{kasa1} is equal to
\begin{equation}
K\big(n,\{d,d+1, \ldots, n-1 \}\big)= \displaystyle\sum_{k\ge 0}{{n-(d-1)k} \choose {k+1}}   \label{KZdsub},
\end{equation}
and the $(n-d)$-complexity  \cite{kasa} is 
\[K\big(n, \{1, 2, \ldots, n-d \}\big) = 2^n - (d-2)\cdot 2^{d-1} - 2,  \textrm{ for } n \ge  2d-2 .\]

\noindent For special cases the following propositions can be easily proved.

\begin{proposition} For $n, d_1\le d_2$ positive integers
\[K\big(n, \{d_1,d_1+1, \ldots, d_2\}  \big) \le n+ \sum_{k\ge 1}{\binom{n-(d_1-1)k}{k+1}} -  \sum_{k\ge 1}{\binom{n-d_2k}{k+1}}.\] 
\end{proposition}

\begin{proof} This can be obtained from (\ref{KZdsub}) and the formula
\begin{eqnarray*}
K\big(n, \{d_1,d_1+1, \ldots, d_2\}  \big) & \le &  K\big(n, \{d_1,d_1+1, \ldots, n-1\}  \big)\\
      &-& K\big(n, \{d_2+1,d_2+2, \ldots, n-1\}  \big) +n. 
\end{eqnarray*}
\end{proof}

For example, $K(7, \{2,3,4,5,6\})=33$, $K(7, \{4,5,6\})=13$, and from the proposition $K(7, \{2,3\})\le 27$. The exact value is $K(7, \{2,3\})=25$, the two words $acg$ and $aeg$ are not eliminated (here the original distances are 2 and 4 in $acg$, and 4 and 2 in $aeg$).

\begin{proposition} For the integers $n,d\ge 1$, where $n=hd+m$
\[K(n, \{d\})=\frac{(h+1)(n+m)}{2}.\]  
\end{proposition}
\begin{proof}
\vspace*{-0.5cm}   
\begin{eqnarray*}
K(n, \{d\})&=& n+ \sum_{i=1}^{n-d}{\left\lfloor \frac{n-i}{d}\right\rfloor=n+d(1+2+\ldots + h-1)+mh}\\
           &=& n+\frac{dh(h-1)}{2}+mh = \frac{(h+1)(n+m)}{2}.
\end{eqnarray*}

\end{proof}
   
To compute the $M$-complexity of a rainbow word of length $n$ we will use graph theoretical results. 
 Let us consider the rainbow word  $a_1a_2\ldots a_n$ and the correspondig digraph $G=(V, E)$, with 

$V=\big\{ a_1, a_2, \ldots, a_n \big\}$, 

$E=\big\{ (a_i,a_j) \mid j-i\in M, \, i=1,2,\ldots, n,  j=1,2,\ldots, n  \big\}$. 

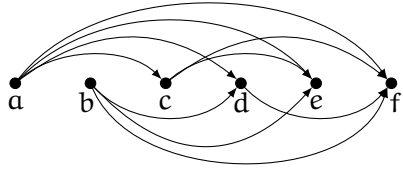
\begin{figure}
\begin{center}
\begin{tikzpicture}
\begin{scope}[>=latex]
\filldraw[]     (1,2) circle (2pt) 
                (2,2) circle (2pt)
                (3,2) circle (2pt)  
                (4,2) circle (2pt)
                (5,2) circle (2pt) 
                (6,2) circle (2pt);
\draw [->] (1,2) .. controls (1.5,2.5) and (2.5,2.5) .. (2.95,2.05);
\draw [->] (1,2) .. controls (2,2.8) and (3,2.8) .. (3.95,2.05);
\draw [->] (1,2) .. controls (2,3.1) and (4,3.1) .. (4.95,2.05);
\draw [->] (1,2) .. controls (2,3.4) and (5,3.4) .. (5.95,2.05);
\draw [->] (3,2) .. controls (3.5,2.5) and (4.5,2.5) .. (4.95,2.05);
\draw [->] (3,2) .. controls (4,2.8) and (5,2.8) .. (5.95,2.05);
\draw [->] (4,2) .. controls (4.5,1.4) and (5.5,1.4) .. (5.95,1.95);
\draw [->] (2,2) .. controls (2.5,1.4) and (3.5,1.4) .. (3.95,1.95);
\draw [->] (2,2) .. controls (3,0.9) and (4,0.9) .. (4.95,1.95);
\draw [->] (2,2) .. controls (2.5,0.6) and (5.5,0.6) .. (5.95,1.95);
\node (1) at (1,1.75) {$a$};
\node (2) at (1.95,1.75) {$b$};
\node (3) at (3,1.75) {$c$};
\node (4) at (4,1.75) {$d$};
\node (5) at (5,1.75) {$e$};
\node (6) at (6.05,1.75) {$f$};
\end{scope}
\end{tikzpicture}
\end{center}\vspace*{-0.8cm}
\caption{Graph for $(2,n-1)$-subwords when $n=6.$}\label{fig1}
\end{figure}

For $n=6, M=\{2,3,4,5\}$ see Figure \ref{fig1}.

The adjacency matrix  $A=\big(a_{ij}\big)_{\tiny i =\overline{1,\!n}, 
                                                            j=\overline{1,\!n}}$ 
of the graph is defined by:

\[a_{ij}=\left\{  \begin{array}{ll}
                    1,  &  \textrm{if } j-i\in M,\\
                    0,     &  \textrm{otherwise},
                 \end{array}  \quad \textrm{  for  }   i=1,2,\ldots, n, j=1,2,\ldots, n.                        
           \right.
\]

Because the graph has no directed cycles, the entry in row $i$ and column $j$ in  $A^k$ (where $A^k=A^{k-1}A$, with $A^1=A$) will represent the number of  directed paths of length $k$ from $a_i$ to $a_j$.  If $I$ is the identity matrix  (with entries equal to 1 only on the first diagonal, and 0 otherwise), let us define the matrix $R= (r_{ij})$:
\[R= I+A+A^2+\cdots + A^k, \textrm{ where } A^{k+1}=O \, (\textrm{the null matrix}).\]
The $M$-complexity of a rainbow word is then
\[ K(n,M)= \sum_{i=1}^{n}{\sum_{j=1}^{n}{r_{ij}}}.\]
Matrix $R$  can be better computed using a variant of the well-known Warshall algorithm (for the original Warshall algorithm see for example \cite{baase}): 
 
\begin{algN}{Warshall($A,n$)}
1 \'  $W \leftarrow A$\\
2 \' \key{for} \= $k \leftarrow 1$ \key{to} $n$ \\
3 \'           \> \key{do}  \key{for} \= $i \leftarrow 1$ \key{to} $n$ \\
4 \'           \>                      \>  \key{do} \key{for} \= $j \leftarrow 1$ \key{to} $n$ \\ 
5 \'            \>                     \>             \>  \key{do} $w_{ij}\leftarrow w_{ij}+w_{ik}w_{kj}$   \\ 
6 \'  \key{return} $W$
\end{algN}

\noindent From $W$ we obtain easily $R=I+W$.

\noindent For example let us consider the graph in Figure \ref{fig1}. The corresponding adjacency matrix is:
\[A=\left(\begin{array}{cccccc}
0&0&1&1&1&1 \\
0&0&0&1&1&1 \\
0&0&0&0&1&1 \\
0&0&0&0&0&1 \\
0&0&0&0&0&0 \\
0&0&0&0&0&0 \\
\end{array}
\right) 
\]
After applying the Warshall algorithm:
\[W=\left(\begin{array}{cccccc}
0&0&1&1&2&3 \\
0&0&0&1&1&2 \\
0&0&0&0&1&1 \\
0&0&0&0&0&1 \\
0&0&0&0&0&0 \\
0&0&0&0&0&0 \\
\end{array}
\right),
\qquad 
R=\left(\begin{array}{cccccc}
1&0&1&1&2&3 \\
0&1&0&1&1&2 \\
0&0&1&0&1&1 \\
0&0&0&1&0&1 \\
0&0&0&0&1&0 \\
0&0&0&0&0&1 \\
\end{array}
\right)
\]
and then $K\big(6,\{2,3,4,5\}\big)=20,$  the sum of elements in $R$.

The Warshall algorithm combined with the Latin square method can be used to obtain all nontrivial (with length at least 2) $M$-subwords of a given rainbow word $a_1a_2\cdots a_n$. Let us consider a matrix ${\cal A}$ with the entries $A_{ij}$, which are set of words. Initially this matrix is defined as:
\[A_{ij}=\left\{  \begin{array}{ll}
                    \{a_ia_j\},  &  \textrm{if } j-i\in M,\\
                    \emptyset,     &  \textrm{otherwise},
                 \end{array}  \quad \textrm{ for } \, i=1,2,\ldots, n, \, j=1,2,\ldots, n. 
           \right.
\]
If ${\cal A}$ and ${\cal B}$ are sets of words, ${\cal AB}$ will be formed by the set of concatenation of each word from ${\cal A}$ with each word from ${\cal B}$: 
\[ {\cal AB} = \big\{ ab  \, \big| \, a\in {\cal A},  b\in {\cal B}  \big\}.
\] 
If $s=s_1s_2\cdots s_p$ is a word, let us denote by $'s$ the word obtained from $s$ by erasing the first character: $'s=s_2s_3\cdots s_p$.  Let us denote by $'{A_{ij}}$ the set ${A_{ij}}$ in which we erase  the first character from each element. In this case $'{\cal A}$ is a matrix with entries $'A_{ij}.$ 

Starting with the matrix ${\cal A}$ defined as before, the algorithm to obtain all nontrivial $M$-subwords is the following:

\begin{algN}{Warshall-Latin(${\cal A},n$)}
1 \'  ${\cal W} \leftarrow {\cal A} $\\
2 \' \key{for} \= $k \leftarrow 1$ \key{to} $n$ \\
3 \'           \> \key{do}  \key{for} \= $i \leftarrow 1$ \key{to} $n$ \\
4 \'           \>                      \>  \key{do} \key{for} \= $j \leftarrow 1$ \key{to} $n$ \\ 
5 \'            \>                     \>             \>  \key{do} \key{if} \= $W_{ik}\ne \emptyset$ and  $W_{kj}\ne \emptyset$ \\
6 \'        \>      \>             \>   \> \key{then} $W_{ij}\leftarrow W_{ij} \cup W_{ik}\, 'W_{kj}$   \\ 
7 \'  \key{return} ${\cal W}$
\end{algN}

The set of nontrivial $M$-subwords is ${\displaystyle\bigcup_{i,j\in \{ 1,2,\ldots, n \} } W_{ij}}$. 

For $n=8$, $M=\{3,4,5,6,7\}$ the initial matrix is:
\begin{center}
$\left(\begin{array}{cccccccc}
\emptyset & \emptyset & \emptyset & \{ad\} & \{ae\} &\{af\} &\{ag\} &\{ah\} \\ 
\emptyset & \emptyset & \emptyset & \emptyset & \{be\}& \{bf\} & \{bg\}& \{bh\} \\ 
\emptyset & \emptyset & \emptyset & \emptyset & \emptyset & \{cf\}&  \{cg\} & \{ch\} \\ 
\emptyset & \emptyset & \emptyset & \emptyset & \emptyset &\emptyset & \{dg\} &\{dh\} \\ 
\emptyset & \emptyset & \emptyset & \emptyset & \emptyset &\emptyset &\emptyset & \{eh\}\\ 
\emptyset & \emptyset & \emptyset & \emptyset & \emptyset &\emptyset &\emptyset & \emptyset \\
\emptyset & \emptyset & \emptyset & \emptyset & \emptyset &\emptyset &\emptyset & \emptyset \\
\emptyset & \emptyset & \emptyset & \emptyset & \emptyset &\emptyset &\emptyset & \emptyset \\
\end{array}
\right).
$
\end{center}

The result of the algorithm \textsc{Warshall-Latin} in this case is: 

\begin{center}
$\left(\begin{array}{cccccccc}
\emptyset & \emptyset & \emptyset & \{ad\} & \{ae\} &\{af\} &\{ag, adg\} &\{ah,adh,aeh\} \\ 
\emptyset & \emptyset & \emptyset & \emptyset & \{be\}& \{bf\} & \{bg\}& \{bh,beh\} \\ 
\emptyset & \emptyset & \emptyset & \emptyset & \emptyset & \{cf\}&  \{cg\} & \{ch\} \\ 
\emptyset & \emptyset & \emptyset & \emptyset & \emptyset &\emptyset & \{dg\} &\{dh\} \\ 
\emptyset & \emptyset & \emptyset & \emptyset & \emptyset &\emptyset &\emptyset & \{eh\}\\ 
\emptyset & \emptyset & \emptyset & \emptyset & \emptyset &\emptyset &\emptyset & \emptyset \\
\emptyset & \emptyset & \emptyset & \emptyset & \emptyset &\emptyset &\emptyset & \emptyset \\
\emptyset & \emptyset & \emptyset & \emptyset & \emptyset &\emptyset &\emptyset & \emptyset \\
\end{array}
\right).
$
\end{center}
 
The algorithm \textsc{Warshall-Latin} can be used for nonrainbow words too, with the remark that repeating subwords must be eliminated. For the word $aabbbaaa$ and $M=\{3,4,5,6,7\}$ the result is: $aa$, $ab$, $aba$, $ba$.

\section{Computing the ($d_1,d_2$)-complexity}

Let us denote by $a_i$ the number of ($d_1,d_2$)-subwords which terminate at position $i$ in a rainbow word of length $n$. Then
\begin{equation}
a_i=1+a_{i-d_1}+ a_{i-d_1-1}+ \cdots + a_{i-d_2},
\label{KZe1} 
\end{equation}
 with the remark that for $i\le 0$ we have $a_i=0$.
Subtracting $a_{i-1}$ from $a_i$ we get the following simpler equation.
\begin{equation}
a_i=a_{i-1}+a_{i-d_1}-a_{i-1-d_2}.
\nonumber 
\end{equation}

The ($d_1,d_2$)-complexity of a rainbow word of length $n$ is 
\begin{equation}
K\big(n, \{d_1,d_1+1, \ldots, d_2\}\big) =   \sum_{i=1}^{n}{a_i}
\label{KZe3}
\end{equation}

For example, if $d_1=2, d_2=4$, the following values are obtained

\medskip\begin{center}
\begin{tabular}{|c||rrrrrrrrrrrrr|} \hline
$n$           & 1 & 2 & 3 & 4 & 5 & 6 & 7  & 8 & 9 & 10 & 11 & 12  & 13 \\ \hline \hline
$a_n$          &1 & 1 & 2 & 3 &  5&  7&  11& 16& 24& 35 &  52&  76 &112 \\ \hline
$K(n,\{2,3,4\})$& 1& 2 & 4 &7  &12 &19 &30  &46 &70 &105 &157 &233  &345 \\ \hline 
\end{tabular}
\end{center}

If we denote by $A(z)=\displaystyle\sum_{n\ge 1}{a_nz^n}$ the generating function of the sequence $a_n$, then from (\ref{KZe1}) we obtain
\[\sum_{n\ge 1}{a_nz^n}=\sum_{n\ge 1}z^n+\sum_{n\ge 1}{a_{n-d_1}z^{n-d_1}}+\cdots +\sum_{n\ge 1}{a_{n-d_2}z^{n-d_2}},\]
and
\[A(z)= \frac{z}{1-z} +z^{d_1}A(z)+\cdots + z^{d_1}A(z). \]
From this we obtain
\begin{equation}
A(z)= \frac{z}{z^{d_2+1}-z^{d_1}-z+1}. \label{Az1}
\end{equation}

For $d_1=2, d_2=4$ the sequence $(a_n)_{n\ge 0}$ (\cite{online} sequence A023435) corresponds to a variant of the dying rabbits problem  \cite{hoggatt}.

To  compute the generating function for the complexity $K\big(n, \{d_1,d_1+1, \ldots,$ $ d_2\}\big)$, let us denote  this complexity simply by $K_n$ only, and its generating function by $K(z)=\displaystyle\sum_{n\ge 1}{K_nz^n}$. We remark that   $K_n=0$  for $n\le 0$, and $K_1=1.$

From (\ref{KZe3}) and (\ref{Az1}) we can immediately conclude that 
\[K(z)=\frac{1}{1-z}A(z)=\frac{z}{(1-z)(z^{d_2+1}-z^{d_1}-z+1)}\, .\]

\section{Correspondence between $(d,n+d-1)$-subwords and $\{1,d\}$-subwords} 
The following result is inspired from the sequence A050228\footnote{A050228: $a_n$ is the number of subsequences $\{s_k\}$ of $\{1,2,3,...n\}$ such that $s_{k+1}-s_k$ is 1 or 3.} of \cite{online}.

\begin{proposition}
The number of $\,\{1,d\}$-subwords of a rainbow word of length $n$ is equal to the number of $\{d,d+1, \ldots, n+d-1\}$-subwords of length at least 2 of a rainbow word of length $n+d$.
\end{proposition}

\begin{proof}
By the generalization of the sequence A050228 \cite{online} the number of the $\{1,d\}$-subwords of a rainbow word of length $n$ is equal to
\[ K\big(n, \{1,d\}\big) = \sum_{k\ge 0}{\binom{n+1-(d-1)k}{k+2}}.\] From (\ref{KZdsub}) we have 
\[K\big(n+d, \{d,d+1,\ldots, n+d-1 \}\big) -(n+d)=  \sum_{k\ge 1}{\binom{n+d-(d-1)k}{k+1}}. \] By changing $k$ to $k+1$ in the sum, we obtain $\displaystyle\sum_{k\ge 0}{\binom{n+1-(d-1)k}{k+2}}$, and this proves the theorem. 
\end{proof}

\noindent\textbf{Example.} For $abcde$ the 19  $\{1,3\}$-subwords are: 
\\
$a,b,c,d,e, ab,  abc, abcd,ad,$ $ ade,  abcde, abe, bc, bcd, bcde,be,
cd, cde, de.$ 

For $abcdefgh$ the 19 $\{3,4,5,6,7\}$-subwords of length at least 2 are: 
\\
$ad, ae, af, ag, adg, ah, adh, aeh, be, bf, bg, bh, beh, cf, cg, ch, dg, dh, eh.$

\section*{Conclusions}
A special scattered subword, the so-called $M$-subword is defined, in which the distances (gaps) between letters are from the set $M$. The number of the $M$-subwords of a given word is the $M$-complexity. Graph algorithms are used to compute the $M$-complexity and to determine all $M$-subwords of a rainbow word. This notion of $M$-complexity is a generalization of the $d$-complexity \cite{ivanyi} and of the super-$d$-complexity \cite{kasa1}. If $M$ consists of successive numbers from $d_1$ to $d_2$ then the so-called $(d_1,d_2)$-complexity is computed by recursive equations and generating functions.

\section*{Acknowledgements}
This work was supported by the project under the grant agreement no. T\'AMOP
4.2.1/B-09/1/KMR-2010-0003 (E\"otv\"os Lor\'and University, Budapest) financed by the European Union and the European Social Fund.

\rightline{\emph{Received:  December 4,  2010 {\tiny \raisebox{2pt}{$\bullet$\!}} Revised: March 12, 2011}}

\end{document}